\documentclass[twocolumn, pra, showpacs,superscriptaddress,floatfix]{revtex4}
\usepackage{graphicx}
\usepackage{dcolumn}
\usepackage{bm}
\usepackage{amsmath}

\begin{document}

\title{Realization of effective super Tonks-Girardeau gases via strongly attractive one-dimensional Fermi gases}

\author{Shu Chen}
\affiliation {Institute of Physics,
Chinese Academy of Sciences, Beijing 100190, China}
\author{Xi-Wen Guan}
\affiliation{Department of Theoretical Physics, Research School of
Physics and Engineering, Australian National University,
Canberra ACT 0200, Australia}
\author{Xiangguo Yin} \affiliation {Institute
of Physics, Chinese Academy of Sciences, Beijing 100190, China}
\author{Liming Guan} \affiliation {Institute of Physics,
Chinese Academy of Sciences, Beijing 100190, China}

\author{M T Batchelor}
\affiliation{Department of Theoretical Physics, Research School of
Physics and Engineering, Australian National University,
Canberra ACT 0200, Australia}
\affiliation{Mathematical Sciences Institute,
Australian National University, Canberra ACT 0200,  Australia}


\begin{abstract}
A significant feature of the one-dimensional super Tonks-Girardeau gas is
its metastable gas-like state with a stronger Fermi-like pressure than for free fermions
which prevents a collapse of atoms.
This naturally suggests a way to search for such
strongly correlated behaviour in systems of interacting fermions in one dimension.
We thus show that the strongly attractive Fermi
gas without polarization can be effectively described by a
super Tonks-Girardeau gas composed of bosonic Fermi pairs with
attractive pair-pair interaction.
A natural description of such super Tonks-Girardeau gases is provided by
Haldane generalized exclusion statistics.
In particular, they are equivalent to ideal
particles obeying more exclusive statistics than Fermi-Dirac statistics.

\end{abstract}

\date{\today}
\maketitle



{\it Introduction.---} Recent experimental progress in manipulating
cold atoms in reduced one-dimensional (1D)  geometry
\cite{gorlitz,esslinger,Paredes,Toshiya}
has stimulated intensive study of the physical properties
of quantum gases, among which an important benchmark is the
experimental realization of
Tonks-Girardeau (TG) gases \cite{Paredes,Toshiya}. For the effective
1D systems,  the effective 1D interactions can be tuned
to reach the strongly interacting regime via Feshbach resonance or
confinement-induced resonance \cite{Olshanii}.
The most recent experimental breakthroughs are the realization of
a 1D super TG (sTG) gas of bosonic Cesium atoms \cite{Haller}
and a 1D spin-imbalanced Fermi gas of $^6$Li atoms \cite{Hulet}.

Whereas the TG gas describes the
strongly repulsive Bose gas \cite{Girardeau,Lieb}, the  sTG gas
describes a gas-like phase of the attractive Bose gas which can be
described by a system of attractive hard rods \cite{Astrakharchik1, Astrakharchik2}.
The sTG gas state corresponds to a highly
excited state in the integrable interacting Bose gas with attractive
interaction \cite{Batchelor}.
Although the sTG state is a highly
excited state which in principle should decay into the cluster ground
state \cite{McGuire2,Tempfli} of the attractive Bose gas, such a state is found to be realized and
stabilized by switching the interactions between bosons from strongly
repulsive to strongly attractive \cite{Haller}.  Due to
the large kinetic energy inherited from the TG phase, the hard core
behavior of the particles with Fermi-like pressure prevents the
collapse of the sTG phase after the switch of interactions from
repulsive to attractive \cite{Batchelor,Astrakharchik1}.

In this work, we propose a scheme to realize the sTG gas in a Fermi
system with attractive interactions. In contrast to the realization
in the attractive interaction regime of the Bose gas \cite{Haller},
the sTG gas is composed of composite bosons which are bound pairs of
fermions with opposite spins and thus is a true ground state (GS).
We further demonstrate that
such a sTG gas is identical to a system of ideal particles
obeying Haldane generalized exclusion statistics (GES)
\cite{Haldane} where the particles and holes are not equally
weighted. In this sense, sTG and Fermi gases  may also provide
insight into the conceptual understanding of Haldane GES, which may
possibly be counted by manipulating ultra cold atoms.

{\it Attractive fermion model.---} We consider a system composed of
two hyperfine components with identical particle numbers
$N_{\uparrow}=N_{\downarrow}=N/2$ in an elongated potential trap with
$\omega _{\perp }\gg \omega _x$, where $\omega_x$ and $\omega _{\perp
}\equiv \omega _y=\omega _z$ are angular frequencies in the axial and
radial directions respectively. $N$ is the total number of fermions.
Under the condition $\omega _{\perp }/\omega _x\gg N$, such Fermi
systems are dynamically described by an effective 1D Hamiltonian
\begin{equation}
H=\sum_{i=1}^N -\frac{\hbar ^2}{2m_F}\frac{\partial ^2}{\partial x_i^2}%
 + g_{F}\sum_{i<j}\delta (x_i-x_j), \label{H1}
\end{equation}
where $g_{F}=-2\hbar ^2/(m_F a^F_{1 \rm D})$ is the effective 1D
interaction strength related to the three-dimensional $s$-wave
scattering length $a^F_s$ by \cite{Olshanii}
$a^F_{1 \rm D}=-l_{\perp }\left( \frac{ l_{\perp }}{a^F_s}-\frac{\left|
\zeta (1/2)\right| }{\sqrt{2}}\right) $ with $ l_{\perp
}=\sqrt{\hbar /m\omega _{\perp }}$ the characteristic oscillator
length in the radial direction.

The eigenvalues of Hamiltonian (\ref{H1}) are given by
$
E=\frac{\hbar^2}{2m_F} \sum_{j=1}^{N} k_j^2
$
with $k_j$ determined by the Bethe ansatz equations (BAE)
\cite{Yang,Gaudin}
\begin{align*}
\exp\left( \mathrm{i} k_{j}L\right) & =\prod_{\alpha=1}^{M} \frac{%
k_{j}-\Lambda_{\alpha}+ \mathrm{i} c_F/2}{k_{j}-\Lambda_{\alpha }- \mathrm{i} c_F/2}%
,\\
\prod_{j=1}^{N}  \frac{\Lambda_{\alpha}-k_{j}+ \mathrm{i} c_F/2}{%
\Lambda_{\alpha}-k_{j}- \mathrm{i} c_F/2} & =-\prod_{\beta=1}^{M}
 \frac{\Lambda_{\alpha}-\Lambda_{\beta}+ \mathrm{i} c_F}{%
\Lambda_{\alpha}-\Lambda_{\beta}- \mathrm{i} c_F} ,
\end{align*}
where $c_F= m_F g_F /\hbar^2 = -2/a^F_{1 \rm D}$, $j=1, \ldots , N$,
$\alpha=1, \ldots, M$ and $M=N/2$ is the number of fermions with
spin down.


This interacting fermion model has been widely studied (see, e.g.,
Refs \cite{Fuchs,Tokatly,Egger05,Wadati,BBGO,orso,hu,Guanxw07} and
references therein).
For strongly attractive interaction, i.e., $L|c_F| \gg
1$, the GS solutions of the BAE correspond to $M=N/2$ pairs of
neutral charge bound states with $k_{\alpha}=\Lambda _{\alpha }\pm
\mathrm{i} c_F /2 + O(\delta)$ for $\alpha=1,\ldots, M$.  Here all
$\Lambda$'s are real and $\delta$ is a very small number of order
$\exp(-L |c_F|)$ \cite{Takahashi}. The BAE thus reduce to
\begin{align}
\exp\left( 2\mathrm{i}\Lambda_{\alpha} L\right) & =-\prod_{\beta=1}^{M} \frac{
\Lambda_{\alpha}-\Lambda_{\beta}+ \mathrm{i} c_F}{\Lambda_{\alpha}-\Lambda_{\beta}-\mathrm{i} c_F}.
\label{Lamda}
\end{align}
The eigenvalues of Hamiltonian
(\ref{H1}) are given by
$ E= -M \epsilon_b+\frac{\hbar^2}{2m_F} \sum_{\alpha=1}^{M} 2
\Lambda_{\alpha}^2 $ where the binding energy $\epsilon_b=({\hbar
^2}/{2m_F}) c_F^{2}/2$, which characterizes internal energy and the
other energy terms include the kinetic energy of the bound pairs and
marginally interacting energy produced from pair-pair scattering in
the strongly attractive interaction limit. In this limit and in the
absence of an external field, we may subtract the binding energy
from the energy, i.e.
\begin{equation}
E_0^{\rm F}=E+M \epsilon_b=\frac{\hbar^2}{2m_F} \sum_{\alpha=1}^{M} 2
\Lambda_{\alpha}^2\label{E-F-e}.
\end{equation}
For strong coupling, the explicit $\Lambda$'s follow from the BAE
(\ref{Lamda}), i.e., $\Lambda_{m} \approx
\frac{(2m+1)\pi}{2L}\left(1-\frac{M}{L|c_F|}\right)^{-1}$ (up to
order $1/c_F^2$), with $m=-M/2,-M/2+1,\ldots, M/2-1$. Here we assume
$M$ is even. The GS energy follows as \cite{BBGO,Wadati}
\begin{eqnarray}
E_0^{\rm F} &\approx& \frac{\hbar ^2}{2m_F} \frac{1}{6}M\left( M^{2}-1\right)
\frac{\pi ^{2}}{L^{2}}\left( 1-\frac{M}{L\left\vert c_F \right\vert
}\right)^{-2}. \label{E0}
\end{eqnarray}

{\it Equivalence to a sTG gas.---}  On the other hand, the 1D interacting Bose gas
composed of $N_B$ bosons is described by the Hamiltonian
\begin{equation}
H=\sum_{i=1}^{N_B} -\frac{\hbar ^2}{2m_B}\frac{\partial ^2}{\partial x_i^2}%
 + g_{B}\sum_{i<j}\delta (x_i-x_j), \label{Hb}
\end{equation}
with $g_{B}=-2\hbar ^2/(m_B a^B_{1d})$.
The energy eigenvalues are given in terms of the quasi-momenta $k_j$ by
\begin{equation}
E=\frac{\hbar^2}{2m_B} \sum_{j=1}^{N_B} k_j^2,\label{E-B}
\end{equation}
which satisfy the BAE \cite{Lieb}
\begin{equation}
\exp\left( \mathrm{i} k_{j}L\right)  = - \prod_{l=1}^{N_B} \frac{
k_{j}-k_{l}+ \mathrm{i} c_B}{k_{j}-k_{l}- \mathrm{i} c_B} , \label{BAEbose}
\end{equation}
with $c_B= m_B g_B /\hbar^2 = -2/a^B_{1 \rm D}$.

\begin{figure}[tbp]
\includegraphics[width=0.95\linewidth]{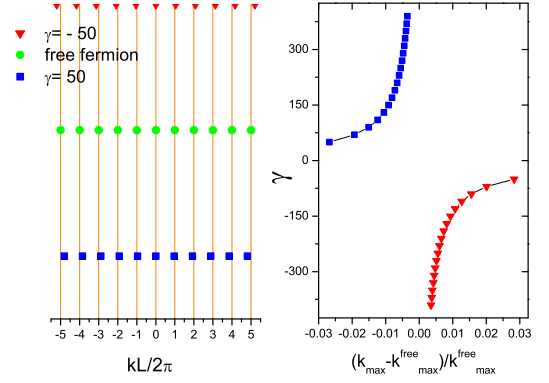}
\caption{ (color online) Quasi-momentum distribution for the GS of
the repulsive Bose gas with $\gamma=50$ and the sTG gas phase
of the attractive Bose gas with $\gamma=-50$ for  $N_B = 11$ (left). The deviation
from the free fermion distribution versus $\gamma$ (right). Here
$\gamma = c_B/n$.} \label{momentum}
\end{figure}

In the TG regime ($c_B \rightarrow \infty$) the quasimomenta
$k_{m}\approx
\frac{(2m+1)\pi}{L}\left(1+\frac{2N_B}{L|c_B|}\right)^{-1}$, with
$m=-N_B/2,-N_B/2+1,\ldots, N_B/2-1$. Here $N_B$ is even. The GS
energy of the strongly repulsive Bose gas in the TG regime (up to
order $1/c_B^2$) is given by
\begin{equation}
E_{TG}  \approx  \frac{\hbar^2}{2m_B} \frac{1}{3} N_B \left(
N_B^{2}-1\right)\frac{\pi ^{2}}{L^{2}}\left( 1 +\frac{2
N_B}{L\left\vert c_B \right\vert }\right)^{-2} . \label{ETG}
\end{equation}

For attractive interaction $c_B<0$, the GS solution for the BAE
(\ref{BAEbose}) is an $N$-string solution and the GS is described by
a cluster state \cite{McGuire2,Tempfli} with energy
$E_0=-\frac{1}{12} c_B^2 N_B(N_B^2-1)$. We note that the BAE
(\ref{BAEbose}) still have real solutions for $c_B<0$, which
obviously correspond to highly excited states. Solving the BAE
(\ref{BAEbose}) gives an explicit form for a gas-like highly excited
state with $k_{m}\approx
\frac{(2m+1)\pi}{L}\left(1-\frac{2N_B}{L|c_B|}\right)^{-1}$, where
$m=-N_B/2,-N_B/2+1,\ldots, N_B/2-1$. Here $N_B$ is even.
In the strongly attractive region ($c_B \rightarrow -\infty$), the
energy of the sTG gas state follows as \cite{Batchelor}
\begin{equation}
E_{STG} \approx \frac{\hbar^2}{2m_B} \frac{1}{3} N_B \left(
N_B^{2}-1\right)\frac{\pi ^{2}}{L^{2}}\left( 1 -\frac{2
N_B}{L\left\vert c_B \right\vert }\right)^{-2} . \label{ESTG}
\end{equation}

Comparing equations (\ref{E0}) and (\ref{ESTG}), it is clear they
are identical if $c_B = 2 c_F$, $N_B=M=N/2$ and
$m_B=2m_F$ (see also Ref. \cite{Wadati}). Since the bound pair formed by two
fermions with opposite spin has a mass $m_B=2m_F$, we can conclude
that the $M$ bound pairs are equivalently described by the sTG
phase of the interacting Bose gas with the effective 1D scattering
length
\begin{equation}
a_{{\rm 1D}}^B=\frac12 {a_{{\rm 1D}}^F}. \label{aa}
\end{equation}
We note that relation (\ref{aa}), obtained by an exact mapping based on
the exact many-body solutions, is consistent with that 
obtained by solving the four-body problem \cite{Egger05}.

The mapping between the GS of the attractive Fermi gas and the
sTG phase of the attractive Bose gas is exact and does not rely
on the strong interaction expansion.  In fact, substituting
$c_F=c_B/2$ into BAE (\ref{Lamda}) and making a replacement $2
\Lambda_{\alpha} =k_{\alpha}$, one finds that BAE (\ref{Lamda}) are
identical to BAE (\ref{BAEbose}) and the energy (\ref{E-F-e}) is
identical to the energy (\ref{E-B}).  To give a concrete example, we
show the solutions of the BAE (\ref{Lamda}) and (\ref{BAEbose}) in
Fig. 1. For $|\gamma|=50$, the roots of
the sTG gas and strongly repulsive Bose gas are very close to
the momentum distributions of free fermions, but on opposite sides
of the free fermion distribution. With $|\gamma| \rightarrow
\infty$, they approach the free fermion distribution.
In Fig. 2, we show the GS energy for the repulsive Bose gas, the
eigenenergy for the sTG gas phase of the attractive Bose gas,
and the GS energy $E_0^F$ for the attractive Fermi gas for different
values of $\gamma$. It is clear that the subtracted GS energy
$E_0^F$ for the attractive Fermi gas is identical to the eigen
energy for the corresponding sTG gas phase with $m_B=2m_F$,
instead of the mass for the Bose gas.

The above conclusion also holds true in the thermodynamic limit
$M,L\rightarrow \infty$ with $n=M/L$ finite, in which the GS energy
(\ref{E-F-e}) can be expressed in the form of the Gaudin integral
equations \cite{Gaudin}.
The Gaudin equations for attractive fermions
coincide exactly with the integral equation form of the sTG
phase -- they do not match the Lieb-Liniger equations for the Bose gas.
We note that this mismatch in the sign of the integral equations
for attractive fermions and repulsive bosons was already noted  \cite{Tokatly}.
However, the ``wrong" sign was argued to be irrelevant in the strong coupling limit.
Here we recognize that the GS of strongly attractive fermions shares the same
signature as the sTG phase of the attractive Bose gas.

The low energy physics of 1D interacting bosons can be described by
Tomonaga-Luttinger liquid (TLL) theory (see \cite{Cazalilla-rev} for
a review).  The TG gas, which describes the strongly repulsive
phase, corresponds to a TLL with $K>1$ ($K\approx (1+2/|\gamma|)^2
$) which characterizes the correlation length, e.g., the one-body
correlation function $g^{(1)}=\langle
\Psi^{\dagger}(x)\Psi(0)\rangle \propto 1/x^{\frac{1}{2K}}$. The sTG
phase corresponds to a highly excited gas-like state where the
particles are strongly correlated. This strongly collective behavior
may be phenomenologically described by the TLL parameter $K \approx
(1-2/|\gamma|)^2$ in the strongly interacting limit
\cite{Batchelor}, which is smaller than $1$.  Consequently, the
paired state of the Fermi gas is also described by a TLL with $K<1$.
In general, a system with $K<1$ sometimes shows CDW quasi-order,
making the system a quasi-supersolid \cite{supersolid}. However, we
notice that the quasi-supersolid phase generally appears in lattice
systems \cite{supersolid,supersolid2} with long range interactions.
For a continuum system with only short range interactions,
the quasi-supersolid phase or CDW order may be hard to realize in general, in
contrast to other ultra-cold atomic systems in optical lattices
\cite{supersolid,supersolid2}.

\begin{figure}[tbp]
\includegraphics[width=0.85\linewidth]{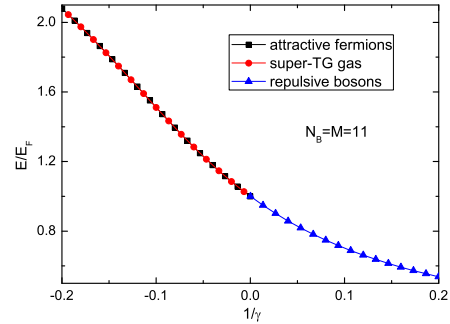}
\caption{ (color online) GS energy for the repulsive Bose gas
(triangles), the eigenenergy for the sTG gas phase of the
attractive Bose gas (dots), and the GS energy $E_0^F$ for the
attractive Fermi gas (squares) versus $1/\gamma$. $E_F$ represents
the GS energy of the TG gas with $\gamma \rightarrow \infty$.}
\end{figure}

{\em Haldane exclusion statistics.---} Cooperative and
collective behavior are significant features of many-body
physics. In 1D pairwise dynamical interaction between identical particles is
inextricably related to their statistical interaction. In particular, coherence
between dynamical interaction and statistical interaction results in
transmutation between these two types of interactions \cite{BG}.  This
can be seen from the equivalence between the 1D Bose gas and Haldane
GES \cite{Haldane}.  This equivalence was set up via an
exact mapping
\begin{equation}
\alpha_{ij}:= \alpha
(k,k')=\delta(k,k')-\frac{1}{2\pi}\theta(k-k'), \label{GES}
\end{equation}
between the Bethe ansatz function  $\theta(k)=2c/(c^2+k^2)$
and the GES parameter $\alpha$ \cite{Wu}.
In general, GES (\ref{GES}) for the 1D Bose gas is mutual
statistics, i.e., $\alpha(k,k')$ depends on all of the other
quasimomenta when moving one particle away from the GS. Importantly,
for the special case of strongly interacting bosons in 1D, Haldane
GES \cite{Haldane} gives a quantitative description of the
fermionization process where the parameter $\alpha_{TG} =
(1+\frac{2N_B}{|c_B|})^{-1}<1$ is nonmutual \cite{BG}. In this case,
the bosons are strongly correlated and behave like identical
particles with GES $\alpha_{TG}$.
Here we further remark that for attractive bosons the GES
description is not valid due to the existence of string solutions to
the BAE. However, we may view all real Bethe ansatz roots as a GES
distribution. In particular, from the set of quasimomenta $\left\{
k_{m}\approx
\frac{(2m+1)\pi}{L}\left(1-\frac{2N_B}{L|c_B|}\right)^{-1}\right\}$
of the sTG state we conceive that the minimum of separation in
momentum space is larger than that of free fermions. In general, the
momentum separation for identical particles with GES is given by
$\Delta k_j\equiv 2\pi(\alpha +\ell)$ \cite{Ha}, where $\ell $ can
be an arbitrary integer. For free fermions the minimum separation of
the momentum is $2\pi/L$ with $\alpha=1$.  This minimum $\alpha$
naturally results in unequal weights for particle density $\rho(k)$
and hole density $\rho_h(k)$ distributions. We understand that for
the sTG gas and the TG gas $\alpha$ number of bosons removed
from the GS creates one hole, i.e.,
\begin{equation}
2\pi \left(\alpha \rho(k)+\rho_h(k)\right)\approx 1,
\end{equation}
with $\alpha =\alpha_{TG}$ or $\alpha_{sTG}$.
This gives the Haldane GES description with nonmutual GES.
In this sense the recent experimental
measurements in a 1D sTG gas of Cesium atoms \cite{Haller} may also provide a
measure of Fermi-like pressure induced from the GES parameter
$\alpha_{STG}=(1-\frac{2N_B}{|c_B|})^{-1}$ which is greater than the pure
Fermi statistics value $\alpha=1$.

For strongly attractive fermions in the absence of an external field, the
neutral charge bound pairs become bosonic hard-core bosons with
nonmutual GES statistics $\alpha_{F}=(1-\frac{M}{|c_F|})^{-1}$
\cite{Guanxw07} . It is clearly seen that the GES parameters
$\alpha_{STG}$ for the sTG gas and $\alpha_{F}$ for bound pairs of
fermions are equivalent under the mapping $c_B = 2 c_F$,
$N_B=M=N/2$. The nonmutual GES for the TG gas, sTG gas and strongly
attractive fermions can be unified  by the most probable distribution
$n(\epsilon)$
\begin{equation}
n(\epsilon)=\frac{1}{\alpha+w(\epsilon)}, \label{GES-n}
\end{equation}
where the function $w(\epsilon)$ satisfies the equation
\begin{equation}
w^{\alpha}(\epsilon)\left(1+w(\epsilon)\right)^{1-\alpha}=e^{\epsilon-\mu/K_BT},
\end{equation}
with $\mu$ the Fermi-like cut-off energy. Here we can easily see
that for $\alpha=0$ and $\alpha=1$ the most probable distribution
$n(\epsilon)$ (\ref{GES-n}) reduces to Bose-Einstein statistics and
Fermi-Dirac statistics, respectively.

Now for TG and sTG bosons we have
$ N_B=\int_{0}^{\infty}d\epsilon\, G_B(\epsilon)n(\epsilon)$ and $
E_B=\int_{0}^{\infty}d\epsilon\, G_B(\epsilon)n(\epsilon)\epsilon$
with density of states
$G_B(\epsilon)=L/\sqrt{2\pi^2\hbar^2\epsilon/m_B}$.
On the other hand, for attractive fermions
$ N_F=2\int_{0}^{\infty}d\epsilon \,G_F(\epsilon)n(\epsilon)$ and
$E_0^F=2\int_{0}^{\infty}d\epsilon\,
G_F(\epsilon)n(\epsilon)\epsilon $ with pair state density
$G_F(\epsilon)=L/\sqrt{\pi^2\hbar^2\epsilon/m_F}$.  For zero
temperature, the GS energies of the TG gas and strongly attractive
fermions are easily obtained through their nonmutual GES
(\ref{GES-n}), along with the excited state energy for the sTG
gas. The sTG gas result (\ref{ESTG}) can also be obtained from
the minima of separation in quasimomentum space derived from GES.
The GES approach provides an alternative way to describe the
thermodynamics of these models.


In summary, we have studied the equivalence
between the GS of the strongly attractive Fermi gas and the sTG gas.
We have shown that Haldane GES provides a natural description
of these strongly correlated states.  By comparing strongly
attractive fermions with the Bose gas, we find that {the bound Fermi
pairs formed in the strongly attractive regime should be described by
the sTG phase of the LL model of attractive bosons, rather than
the LL model of repulsive bosons.} This finding suggests that we can
realize the sTG gas by preparing a 1D Fermi gas in the strongly
attractive regime.  Since the Fermi pairs are unbreakable in the
strongly attractive limit, such a state is expected to be very
stable.
Moreover, our results suggest that experimental observation of Haldane
statistics can be done by detecting the breathing mode of
the attractive Fermi gas without polarization and comparing with the
result obtained from the integrable anyon model with GES
parameter $\alpha$ as fitting parameter \cite{BG}.

{\it Acknowledgments.---} SC has been supported by the NSF of China
under grants 10821403 and 10974234, 973 grant 2010CB922904.
and the National Program for Basic Research of MOST.
XWG and MTB have been supported by the Australian Research Council.

\end{document}